\documentclass[12pt]{article}
\begin{document}

\title{Noncommutativity and relativity}
\author{B.G. Sidharth\footnote{birlasc@gmail.com}, B.M. Birla Science Centre,\\ Adarsh Nagar, Hyderabad - 500 063, India\\
\\Abhishek Das\footnote{parbihtih3@gmail.com}, B.M. Birla Science Centre,\\ Adarsh Nagar, Hyderabad - 500 063, India\\}
\date{}
\maketitle
\begin{abstract}
In this paper, we endeavour to show that from the noncommutative nature of spacetime one can deduce the concept of relativity in the sense that the velocity cannot be infinite as in the case of Galilean relativity.
\end{abstract}
\maketitle

\section{Introduction}
The noncommutative nature of space-time has been a topic of great
interest and has been studied extensively by several authors
\cite{Michael,Louis}. On the other hand, special relativity is a physical theory apposite to the relationship between space and time, which tells us that space and time are interwoven into a single continuum defined as spacetime. Special relativity implies a wide range of consequences including length contraction, time dilation, relativistic mass, a universal speed limit and so on. In this paper we strive to show that one can extract the relativistic nature of spacetime from noncommutativity itself. Our work is based on Snyder's \cite{Snyder} noncommutative geometry, in contradistinction to the Weyl-Moyal formalism \cite{Chaichian,Raul}.\\
In the first section, starting from basic relations of noncommutativity we show that there is a maximum possible velocity in the universe. From this result the concept of relativity arises. After that, in the subsequent sections we show length contraction, time dilation and relativistic mass arising from the maximal universal velocity. Lastly, we discuss how the results of this paper are connected to special relativity.\\

\section{Noncommutativity and boundedness of velocity}

Previously in a paper, the author Sidharth \cite{bgs1} had shown that along with space noncommutativity one can obtain momentum noncommutativity. This was rigorously substantiated recently \cite{bgs3} by the authors of this paper, such that\\
\begin{equation}
[p_{x}, p_{y}] = \eta\theta(l^{2}) \label{a}
\end{equation}
where, $\eta$ is generally takes the value of '$\pm i$' as considered by various authors \cite{Chaichian,Girotti} and $\theta(l^{2})$ is a $2 \times 2$ matrix where $l$ is the fundamental minimum length usually taken to be the Planck length or the Compton length. We represent $\theta(l^{2})$ in the following manner\\
\[\theta(l^{2}) = f(l^{2})\theta_{xy}\]\\
where, $f(l^{2})$ is a positive, finite and real valued scalar and $\theta_{xy}$ is a $2 \times 2$ matrix. Next, we consider the following representation for the momenta\\
\[p_{x} = P_{x}\sigma_{x}\]\\
and\\
\[p_{y} = P_{y}\sigma_{y}\]\\
where, $P_{x}$ and $P_{y}$ are the scalar values of the momenta and the $\sigma$'s are the Pauli matrices. Thus we have from (\ref{a})\\
\[(P_{x}\sigma_{x})(P_{y}\sigma_{y}) - (P_{y}\sigma_{y})(P_{x}\sigma_{x}) = \eta f(l^{2})\theta_{xy}\]\\
\[P_{x}P_{y}[\sigma_{x}\sigma_{y} - \sigma_{y}\sigma_{x}] = \eta f(l^{2})\theta_{xy}\]\\
which gives
\begin{equation}
P_{x}P_{y}[\sigma_{x}, \sigma_{y}] = \eta f(l^{2})\theta_{xy} \label{b}
\end{equation}
Now, considering the commutation relation of the Pauli matrices given by\\
\[[\sigma_{x}, \sigma_{y}] = 2i\epsilon_{xyz}\sigma_{z}\] \\
we can write from (\ref{b})\\
\begin{equation}
P_{a}P_{b}[\epsilon_{abc}\sigma_{c}] = -\frac{i}{2}\eta f(l^{2})\theta_{ab} \label{f}
\end{equation}
where, $a, b, c$ can take values $x, y, z$ but $a \neq b \neq c$ since the Levi-Civita tensor has to be non-zero in this case, otherwise relation (\ref{f}) will be trivial. Now, for even and odd permutations of $a, b$ and $c$, the sign of the Levi-Civita tensor is adjusted with the sign of $\eta = \pm i$. Therefore, we have\\
\begin{equation}
(P_{a}P_{b})\sigma_{c} = \frac{1}{2} f(l^{2})\theta_{ab} \label{c}
\end{equation}
Now, multiplying both sides of (\ref{c}) by $\sigma_{c}$ we have\\
\[(P_{a}P_{b})\sigma_{c}^{2} = f(l^{2})\theta_{ab}\sigma_{c}\]\\
Now, using the relation\\
\[\sigma_{c}^{2} = I\]\\
we have\\
\begin{equation}\\
(P_{a}P_{b})I = f(l^{2})\beta_{ab} \label{e}
\end{equation}
where, $I$ is the identity matrix and $\beta_{ab} = \theta_{ab}\sigma_{c}$ is another $2 \times 2$ matrix. Now, since relation (\ref{e}) is an equality, taking the determinant of both sides will also yield an equality, such that\\
\begin{equation}
|(P_{a}P_{b})I| = |f(l^{2})\beta_{ab}|
\end{equation}
Since, on both sides we have scalars multiplied to the $2 \times 2$ matrices, we can write using the property of determinants\\
\[(P_{a}P_{b})^{2}|I| = \{f(l^{2})\}^{2}|\beta_{\mu\nu}|\]\\
which gives\\
\[(P_{a}P_{b})^{2} = \{f(l^{2})\}^{2}\epsilon^{2}\] \\
where, \\
\[|\beta_{ab}| = |\theta_{ab}\sigma_{c}| = \epsilon^{2}\]\\
and the matrix $\theta_{ab}$ is such that $\epsilon$ is finite and real. Now, since product of the scalar values of the momenta has to be positive, we finally have\\
\begin{equation}
P_{a}P_{b} = \epsilon f(l^{2}) \label{d}
\end{equation}
Again, the momentum is related to the velocity as\\
\[P = mv\]\\
we can say from (\ref{d}) that\\
\[v_{a}v_{b} = \frac{\epsilon f(l^{2})}{m^{2}}\]\\
Again, $f(l^{2})$ being a finite valued function of the square of the fundamental length is independent of the momentum and the velocity. Therefore, we can say that the product $v_{a}v_{b}$ is bounded. This is because, one can always find a function $g(l^{2})$ (finite and real valued) such that \\
\[g(l^{2}) > f(l^{2})\]\\
Therefore, we have\\
\[v_{a}v_{b} < \frac{\epsilon g(l^{2})}{m^{2}}\]\\
Consequently, the individual velocities $v_{a}$ and $v_{b}$ will be bounded above. From this we can conclude that the velocity of a particle has an upper bound, i.e.\\
\begin{equation}
v \leq \alpha \label{g}
\end{equation}
where, $\alpha$ is some finite velocity and it is the maximum possible velocity in the universe.

\section{Composition of velocities and relativity}
Now, let us consider the following case where there are two frames of references, namely $S$ and $S^{\prime}$. The references frame $S$ is at rest and $S^{\prime}$ is moving with velocity $\omega$ with respect to $S$. Also, a body is moving in the same direction along with the reference frame $S^{\prime}$ and an observer in $S^{\prime}$ measures it's velocity as $\omega^{\prime}$. Therefore, the velocity of the moving body as measured by an observer in $S$ would be given by\\
\begin{equation}
V = \omega + \omega^{\prime}
\end{equation}
Now, according to the relation (\ref{g}) we must have the following\\
\[\omega \leq \alpha\]\\
\[\omega^{\prime} \leq \alpha\]\\
and\\
\[V = \omega + \omega^{\prime} \leq \alpha\]\\
As an example, suppose the values of $\omega$ and $\omega^{\prime}$ are such that\\
\[\omega = \omega^{\prime} \approx 0.99999\alpha\]\\
Ostensibly, the values of the velocities are valid since both $\omega$ and $\omega^{\prime}$ are in agreement with relation (\ref{g}). But, incidentally we have\\
\begin{equation}
V = \omega + \omega^{\prime} = 1.99998\alpha
\end{equation}
which violates the bound given by (\ref{g}), since $\alpha$ is the maximum possible velocity in the universe. So, there must have been a discrepancy somewhere, though every equation is correct. Actually, this problem can be eradicated if we simple consider a factor $\rho$ ($< 1$) such that we have\\
\begin{equation}
V = \rho(\omega + \omega^{\prime}) \leq \alpha \label{k}
\end{equation}
As we can see, there is a divergence from the Galilean relativity of Euclidean geometry. In fact, from here emerges the notion of relativity because of the factor $\rho$ that adjusts the discrepancy that surfaces on account of considering velocities that are nearly equal to $\alpha$. Essentially, the factor $\rho$ depends on the velocities $\omega$, $\omega^{\prime}$ and $\alpha$. Of course, in non-relativistic cases, we would have $\rho = 1$  and get back to Galilean relativity and simple addition of velocities.\\
This is simply the relativistic addition of velocities. Thus, using the noncommutative feature of spacetime one can understand the affine connection between the Minkowski spacetime and relativity itself.\\

\section{Length contraction}

Now, suppose a rod is placed along the $x^{\prime}$ axis of the reference frame $S^{\prime}$ which moves with velocity $v$ with respect to the rest frame $S$. Now, the length of the rod as measured by an observer in $S^{\prime}$ is\\
\[l_{0} = x_{2}^{\prime} - x_{1}^{\prime}\]\\
where, $x_{2}^{\prime}$ and $x_{1}^{\prime}$ are the endpoints of the rod with respect to $S^{\prime}$. Now, an observer in the reference frame $S$ will measure this as\\
\[l = x_{2} - x_{1}\]\\
where, $x_{2}$ and $x_{1}$ are the coordinates of the extremities of the rod with respect to $S$. We will have \\
\begin{equation}
x_{2}^{\prime} - x_{1}^{\prime} = x_{2} - x_{1} \label{m}
\end{equation}
Now, if the measurements of the endpoints are taken accordingly at instant $t_{1}$ and $t_{2}$ then one can say that the velocity of the rod (incidentally, the velocity of $S^{\prime}$ with respect to $S$), $v$, will be given by\\
\[v = \frac{x_{2}^{\prime} - x_{1}^{\prime}}{t_{2} - t_{1}}\]
Again, we also have for the frame $S$\\
\[\omega = \frac{x_{2} - x_{1}}{t_{2} - t_{1}}\]\\
so that\\
\begin{equation}
v = \omega \label{n}
\end{equation}
Now, suppose the velocity increases by an infinitesimal amount $u$, so that we have\\
\begin{equation}
v + u = \omega + u \label{l}
\end{equation}
Here, the velocity '$u$' is very small and negligible ($u \approx 0$). Now, since, $\omega + u > \omega$, and taking into account the fact that in the relation (\ref{k}), due to the boundedness of velocity we had multiplied a factor $\rho$ to obtain\\
\[\rho(\omega + \omega^{\prime})\]\\
instead of $(\omega + \omega^{\prime})$, so in this case we write\\
\begin{equation}
\omega + u = \theta\omega
\end{equation}
where, $\theta > 1$, since $u$ is an arbitrarily small {\it increment} in velocity. Thus, the right hand side of (\ref{l}) must be rewritten as\\
\[v + u = \theta\omega\]\\
so that relation (\ref{m}) and (\ref{n}) are no longer true due to the factor $\theta$. From here, we have\\
\[\frac{x_{2}^{\prime} - x_{1}^{\prime}}{t_{2} - t_{1}} + u = \theta\frac{x_{2} - x_{1}}{t_{2} - t_{1}}\]\\
\[x_{2}^{\prime} - x_{1}^{\prime} + u(t_{2} - t_{1}) = \theta (x_{2} - x_{1})\]\\
Again, the measurement for the length of the rod must be simultaneous in the frame $S$. Suppose, the measurements are not exactly simultaneous and there is an infinitesimal difference such that\\
\[t_{2} - t_{1}\]\\
is very small. Also, $u$ is arbitrarily very small. Thus, the term $u(t_{2} - t_{1})$ can be neglected. Consequently, we have finally\\
\begin{equation}
l_{0} = \theta l \label{q}
\end{equation}
where, $\theta > 1$ from our considerations. This is the result for length contraction.\\

\section{Time dilation}

As in the previous cases, let us consider a rest frame $S$ and frame $S^{\prime}$ moving with velocity $v$ with respect to $S$. Now, suppose a clock is placed at the coordinate $x^{\prime}$ of the frame $S^{\prime}$. So, the distance of the clock from the origin ($O^{\prime}$) of $S^{\prime}$ is $x^{\prime}$. Consider a body situated at the origin $O^{\prime}$ at the time $t_{1}^{\prime}$, recorded by the clock. When the frame $S^{\prime}$ has moved a distance $x^{\prime}$, the body at origin will have covered the distance $x^{\prime}$ arriving at the coordinate $x^{\prime}$. Let the clock measure this instant of the body arriving at the coordinate $x^{\prime}$ be written as $t_{2}^{\prime}$. Therefore, we have\\
\[t_{0} = t_{2}^{\prime} - t_{1}^{\prime} = \frac{x^{\prime}}{v}\]\\
Again, for a clock in the reference frame we should have the time difference as\\
\[t = t_{2} - t_{1} = \frac{x}{\omega}\]\\
such that\\
\begin{equation}
vt_{0} = \omega t \label{p}
\end{equation}
Now, in the previous section we have seen that due to our considerations \\
\[v \neq \omega\]\\
In fact, we have\\
\[v + u = \theta\omega\]\\
where, $\theta > 1$. But, since we had considered $u$ to be an arbitrarily small amount of velocity, we can neglect $u$ and simply write\\
\[v = \theta\omega\]\\
Using this in (\ref{p}) we get\\
\[\theta\omega t_{0} = \omega t\]\\
Since, $\omega \neq 0$, we finally obtain\\
\begin{equation}
t = \theta t_{0} \label{r}
\end{equation}
This is essentially the result for time dilation that is the characteristic attribute of special relativity. Therefore, we can now firmly conclude that the noncommutative nature of spacetime leads to the Minkowski spacetime and thereby gives the results of special relativity. Basically, due to the noncommutative feature we see that space and time are already distorted. Essentially, we can conclude that relativity is borne out of the {\it noncommutative Minkowski spacetime}.\\

\section{Relativistic mass}

Now, let us again consider the reference frames $S$ and $S^{\prime}$ where $S^{\prime}$ moves with velocity $v$ with respect to $S$. Let us also consider two bodies each of mass $m^{\prime}$ moving in opposite directions along the $x^{\prime}$ - axis of the frame $S^{\prime}$. Let, $\omega^{\prime}$ and $-\omega^{\prime}$ be their velocities along the $x^{\prime}$ -axis, according to an observer in $S^{\prime}$. If the bodies coalesce and form a single body then the new body will be at rest according to the conservation law of momentum, with respect to the frame $S^{\prime}$.\\
Now, the velocities of the two bodies according to an observer in the frame $S$ will be given by \\
\[\omega_{1} = \rho_{1}(\omega^{\prime} + v)\]\\
and \\
\[\omega_{2} = \rho_{2}(-\omega^{\prime} + v)\]\\
according to the relation (\ref{k}), where $\omega_{1}$ and $\omega_{2}$ are the velocities along $x$ - axis. Here, $\rho_{1}, \rho_{2} < 1$. Now, suppose $m_{1}$ and $m_{2}$ are the masses of the two bodies according to an observer in the $S$ frame. Then, the body formed when these two individual bodies coalesce will have the mass ($m_{1} + m_{2}$) moving with a velocity $v$, according to the observer in $S$. Thus, due to the conservation law of momentum we will have\\
\[m_{1}\omega_{1} + m_{2}\omega_{2} = (m_{1} + m_{2})v\]\\
Using the aforementioned relations for $\omega_{1}$ and $\omega_{2}$ we have\\
\[m_{1}[\rho_{1}(\omega^{\prime} + v)] + m_{2}[\rho_{2}(-\omega^{\prime} + v)] = (m_{1} + m_{2})v\]\\
Now, dividing both sides of the equation by $m_{2}$ we get\\
\[\frac{m_{1}}{m_{2}}[\rho_{1}(\omega^{\prime} + v)] + \rho_{2}(-\omega^{\prime} + v) = v\frac{m_{1}}{m_{2}} + v\]\\
From this we may rearrange and write\\
\begin{equation}
\frac{m_{1}}{m_{2}} = \frac{v + (\omega^{\prime} - v)\rho_{2}}{\rho_{1}(\omega^{\prime} + v) - v} \label{s}
\end{equation}
Now, without relativistic considerations we should have\\
\[m_{1} = m_{2}\]\\
But, we shall see that it is not possible. Since, the relation (\ref{s}) is arbitrary, so if in case $m_{1} = m_{2}$ we must have\\
\[v + (\omega^{\prime} - v)\rho_{2} = \rho_{1}(\omega^{\prime} + v) - v\]\\
for all values of $\omega^{\prime}$, $v$, $\rho_{1}$ and $\rho_{2}$. Since this is a general relation, if it is invalid for some specific condition then it has to be invalid in general. Let us see. Rearranging the terms we may write\\
\begin{equation}
v[2 - (\rho_{1} + \rho_{2})] = \omega^{\prime}(\rho_{1} - \rho_{2}) \label{v}
\end{equation}
Now, since $\rho_{1}, \rho_{2} < 1$ the left hand side is positive. But, in case of the right hand side it is negative when $\rho_{2} > \rho_{1}$. This is a condition that violates the general relation (\ref{v}). Thus, it cannot be true. Therefore, we may write that\\
\[m_{1} \neq m_{2}\]\\
for relativistic cases. In non-relativistic cases, where $\rho_{1}, \rho_{2} \approx 1$ we would obviously have \\
\[m_{1} = m_{2}\]\\
Thus, the relation (\ref{s}) can be written as\\
\begin{equation}
\frac{m_{1}}{m_{2}} = \zeta
\end{equation}
Therefore, if the velocity of the second body with respect to the system $S$ is observed as zero, i.e. if we have the velocity $\omega_{2} = 0$ then it's mass will be the rest mass given as\\
\[m_{2} = m_{0}\]\\
Now, writing $m_{1} = m$, we finally have\\
\begin{equation}
m = \zeta m_{0} \label{u}
\end{equation}
where, $m$ is the relativistic mass and $\zeta$ ostensibly depends on the velocity $v$ of the moving frame $S^{\prime}$. Now, let us consider the kinetic energy ($T$) of a moving body. It can be written as\\
\[T = \int_{0}^{t} F{\rm d}r\]\\
where, $F$ is generally the component of the force to the displacement ${\rm d}r$. This can also be written as\\
\[T = \int_{0}^{t} \frac{({\rm d}mv)}{{\rm d}t}{\rm d}r\]\\
\[T = \int_{0}^{v} v{\rm d}(\zeta m_{0}v)\]\\
If we integrate this equation by parts then we will obtain\\
\begin{equation}
T = \zeta m_{0}v^{2} - m_{0}\int_{0}^{v} \zeta v{\rm d}v \label{t}
\end{equation}
Now, the to terms on the right hand side of (\ref{t}) shall give the total energy and the rest mass of the body. Therefore, as we can see from our considerations we have been able to derive a relation for the relativistic mass as in (\ref{u}) and a relation for the energy of a moving body as in (\ref{t}).\\

\section{Discussions}

1)~In this paper, we have shown that beginning with the basic considerations of noncommutativity one can extract the results of special relativity. Let us consider, for example the relation (\ref{k})\\
\[V = \rho(\omega + \omega^{\prime}) \leq \alpha\]\\
Here, the factor $\rho$ ostensibly depends on the velocities $\omega$, $\omega^{\prime}$ and $\alpha$ too, in order to satisfy the condition (\ref{g}). Incidentally, taking the cue from special relativity we write\\
\[\rho = (1 + \frac{\omega\omega^{\prime}}{\alpha^{2}})^{-1}\]\\
then we have the known formula arising from special relativity, considering\\
\[\alpha = c\]\\
i.e. the velocity of light. Similarly, in the relation (\ref{q}), i.e.\\
\[l_{0} = \theta l\]\\
the factor $\theta$ depends on the velocity $v$ of the moving reference frame $S^{\prime}$ and will also depend on the limiting velocity $\alpha$. We can write \\
\[\theta = (1 - \frac{v^{2}}{\alpha^{2}})^{-\frac{1}{2}}\]\\
which gives the Lorentz factor when we consider, $\alpha = c$. The same factor $\theta$ emerges again in relation (\ref{r})\\
\[t = \theta t_{0}\]\\
and we have the known result for time dilation. Again, in case of relativistic mass we had got the relation (\ref{v}) where we argued that it is an invalid equation. This can be easily shown when we put $\rho_{1} = \{1 + \frac{\omega^{\prime}v}{\alpha^{2}}\}^{-1}$ and $\rho_{2} = \{1 + \frac{(-\omega^{\prime})v}{\alpha^{2}}\}^{-1}$ since, the bodies have velocities $\omega^{\prime}$ and $-\omega^{\prime}$. It is obvious from these expressions of $\rho_{1}$ and $\rho_{2}$ that\\
\[\rho_{2} > \rho_{1}\]\\
Therefore, the relation (\ref{v}) is invalid in relativistic cases. Now, considering these values of $\rho_{1}$ and $\rho_{2}$ we would have\\
\[\zeta = \frac{\rho_{2}}{\rho_{1}} = \frac{1 + \frac{\omega^{\prime}v}{\alpha^{2}}}{1 + \frac{(-\omega^{\prime})v}{\alpha^{2}}} = \frac{1}{\sqrt{1 - \frac{v^{2}}{\alpha^{2}}}}\]\\
Again, setting $\alpha = c$, we have $\zeta$ as the Lorentz factor. Now, in equation (\ref{t}) we had found\\
\[T = \zeta m_{0}v^{2} - m_{0}\int_{0}^{v} \zeta v{\rm d}v\]\\
Thus, we have\\
\[T = \frac{m_{0}v^{2}}{\sqrt{1 - \frac{v^{2}}{c^{2}}}} - m_{0}\int_{0}^{v} \frac{v}{\sqrt{1 - \frac{v^{2}}{c^{2}}}}{\rm d}v\]\\
After integrating the second term one has\\
\[T = \frac{m_{0}v^{2}}{\sqrt{1 - \frac{v^{2}}{c^{2}}}} + m_{0}c^{2}[\sqrt{1 - \frac{v^{2}}{c^{2}}} - 1]\]\\
which finally gives\\
\[T = mc^{2} - m_{0}c^{2}\]\\
which is the desired result. Thus, it is conspicuous that special relativity can be obtained from the noncommutative nature of spacetime. \\
\\
2)~Now, the infinitesimal increment in velocity ($u$) which we had considered in relation (\ref{l}) can also be looked upon as a virtual increase in velocity. Manifestly, this means that '$u$' leaves the physics of the system of reference frames unchanged in the long run. This infinitesimal velocity was introduced for heuristic purpose and it has no physical significance altogether.

\end{document}